\documentclass[runningheads]{llncs}
\usepackage{graphicx}

\usepackage{amsmath,amssymb,amsfonts,bm}

\usepackage{color}
\def\mycolor{black}
\def\mycolorSec{black}

\begin{document}
\title{Effects of VLSI Circuit Constraints on Temporal-Coding Multilayer Spiking Neural Networks \thanks{partially supported by the ``Brain-Morphic AI to Resolve Social Issues'' project at UTokyo, the NEC Corporation, \textcolor{\mycolorSec}{AMED under Grant Number JP20dm0307009, and JST Moonshot R\&D Grant Number JPMJMS2021.} }}
\titlerunning{Effects of VLSI Circuit Constraints}
%
\author{Yusuke Sakemi\inst{1}\orcidID{0000-0002-0274-9491} \and
Takashi Morie \inst{2}\orcidID{0000-0003-2708-4307} \and
Takeo hosomi\inst{1} \orcidID{0000-0001-5972-3877} \and
Kazuyuki Aihara\inst{3,4}\orcidID{0000-0002-4602-9816}}
\authorrunning{Y. Sakemi et al.}
%
\institute{NEC Corporation, Kawasaki, Japan 
\email{ysakemi@nec.com} \and
Graduate School of Life Science and Systems Engineering, Kyushu Institute of Technology, Kitakyushu, Japan \and
Institute of Industrial Science, The University of Tokyo, Tokyo, Japan \and
International Research Center for Neurointelligence (WPI-IRCN), 
The University of Tokyo Institutes for Advanced Study, 
The University of Tokyo, 
Tokyo, Japan
}
\maketitle              
\begin{abstract}
The spiking neural network (SNN) has been attracting considerable attention not only as a mathematical model for the brain, but also as an energy-efficient information processing model for real-world applications.
\textcolor{\mycolor}{In particular,  SNNs based on} temporal coding are expected to be much more efficient than those \textcolor{\mycolor}{based on} rate coding, 
because the former requires substantially fewer spikes to carry out tasks.
As SNNs are continuous-state and continuous-time models, it is favorable to implement them with analog VLSI circuits. 
However, the construction of the entire system with continuous-time analog circuits \textcolor{\mycolorSec}{would be infeasible when the system size is very large.}  
Therefore, mixed-signal circuits must be employed, 
and the time discretization and quantization of the synaptic weights are necessary.
Moreover, the analog VLSI implementation of SNNs exhibits non-idealities, such as the effects of noise and 
device mismatches, as well as other constraints arising from the analog circuit operation.
In this study, we investigated the effects of the time discretization and/or weight quantization on the performance of SNNs. 
\textcolor{\mycolor}{Furthermore,} we elucidated the effects the lower bound of the membrane potentials and the \textcolor{\mycolorSec}{temporal fluctuation of} the firing threshold.
Finally, we propose an optimal approach for the mapping of mathematical SNN models to analog circuits with discretized time.

\keywords{spiking neural network \and temporal coding \and supervised learning \and VLSI.}
\end{abstract}
\section{Introduction}
Spiking neural networks (SNNs) are expected to be energy-efficient models 
because they can process information in the form of spikes in an event-driven manner.
Moreover, the performance of SNNs has been demonstrated to be comparable to that of artificial neural networks (ANNs) with deep learning in relatively small-scale image recognition tasks~\cite{Tavanaei2019Deep,Pfeiffer2018Deep}. 
The information representations in SNNs can be approximately classified into two domains: rate and temporal coding.
The information is contained in the rate of the spikes in rate coding, whereas
it is contained in the precise timing of the spikes in temporal coding. 
Although rate coding is closely linked to ANNs \cite{Diehl2015fast}\cite{Kim2020spiking} 
in which the neuron states are represented by analog values,
temporal coding can efficiently exploit the information of the temporal dynamics of the membrane potentials and synaptic currents, which are not considered in ANNs~\cite{Huh2018gradient}. 
In particular, time-to-first-spike (TTFS) coding, which is a type of temporal coding, is expected to be the most efficient because the information of a neuron output is represented as the timing of a single spike~\cite{Bohte2002error}\cite{Mostafa2018supervised}\cite{Comsa2019temporal}\cite{Sakemi2020supervised}. 

As SNNs are expressed as continuous-state and continuous-time systems, analog VLSI implementation is favorable.
However, the construction of the entire system, including the input/output parts, \textcolor{\mycolorSec}{would be infeasible} with continuous-time analog circuits \textcolor{\mycolorSec}{when the system size is very large} because of inflexibility in system design and difficulty in analog circuit design. 
The discretization of the neuron activation in the time domain and quantization of the synaptic weights are required. 
Therefore, many studies on SNNs have adopted fully digital VLSI implementation~\cite{Merolla2014}\cite{Davies2018loihi} or mixed-signal VLSI implementation~\cite{Moradi2018scalable}.

In this study, we temporally discretized SNNs trained with TTFS coding and quantized the synaptic weights to facilitate VLSI implementation. 
Moreover, we investigated the effects of the lower bound of the membrane potentials and \textcolor{\mycolorSec}{the temporal fluctuation of} the firing thresholds, which are problematic in analog VLSI implementation. 
Section~\ref{sec:related_works} introduces related works,
Section~\ref{sec:circuit_constraints} presents the investigation into the above-mentioned effects,
and Section~\ref{sec:mapping} demonstrates the optimal mapping of SNNs to analog \textcolor{\mycolor}{VLSI} circuits.

\section{Related works} \label{sec:related_works}

Several supervised learning algorithms for multilayer SNNs \textcolor{\mycolor}{based on temporal coding} have been proposed in discrete time systems~\cite{Kheradpisheh2019s4nn}\cite{Kim2020unifying}.
However, these algorithms require approximations to obtain the derivative of the spike timing with respect to the weights to \textcolor{\mycolor}{derive} the backpropagation algorithms. 
In our approach, the SNNs are trained in continuous time and then discretized, which does not require any approximation to derive the backpropagation algorithm apart from the spike vanishing problem \cite{Bohte2002error}, 
and therefore, the training of the SNNs is easier.
Moreover, as we demonstrate later, the discretized time step for the spike \textcolor{\mycolorSec}{timing} is important to map the SNNs to \textcolor{\mycolor}{VLSI} circuits optimally.
In our method, once the SNNs are trained in a continuous-time system, an arbitrary time step can be selected for the spike \textcolor{\mycolorSec}{timing} without retraining, whereas in the other approaches, the optimization of the time step is difficult because the time step must be fixed when training SNNs~\cite{Kheradpisheh2019s4nn}\cite{Kim2020unifying}. 
Furthermore, we note that it is common to quantize the activation of neurons after training in ANNs~\cite{Sze2017}.

Certain research groups have designed VLSI circuits for SNNs based on TTFS coding~\cite{Goltz2020fast}\cite{Oh2020hardware}.
G{\"o}ltz {\it et al.} designed a mixed-signal circuit and demonstrated SNNs based on TTFS coding~\cite{Goltz2020fast}.
Oh {\it et al.} designed analog SNN circuits with analog memory based on TTFS coding~\cite{Oh2020hardware}.
\textcolor{\mycolor}{Although} Kheradpishesh {\it et al.}~\cite{Kheradpisheh2020bs4nn} and Sakemi {\it et al.} \cite{Sakemi2020supervised} investigated the effects of the timing jitter of spikes on the performance, 
the effects of the time discretization of the spikes have not been discussed. 
Sakemi {\it et al.} \cite{Sakemi2020supervised} and Oh {\it et al.} \cite{Oh2020hardware} investigated the effects of fixed scattering of the firing thresholds owing to device mismatches. 
However, they did not discuss the effects of the \textcolor{\mycolorSec}{temporal fluctuation of} the firing thresholds.
Oh {\it et al.} \cite{Oh2020hardware} investigated the effects of the variations in the synaptic weights,
and Kheradpishesh {\it et al.}~\cite{Kheradpisheh2020bs4nn} binarized weights of SNNs on TTFS coding.
\textcolor{\mycolor}{However,} they did not discuss the effects of weight quantization. 
To the best of our knowledge, the effects of the lower bound of the membrane potentials for SNNs based on TTFS coding have not been investigated in previous works~\cite{Bohte2002error}\cite{Mostafa2018supervised}\cite{Comsa2019temporal}\cite{Sakemi2020supervised}\cite{Goltz2020fast}\cite{Oh2020hardware}. 

\section{Effects of non-idealities arising from VLSI circuit constraints} \label{sec:circuit_constraints}

\subsection{Models}

\begin{figure}
\begin{center}
\includegraphics[clip, width=12cm]{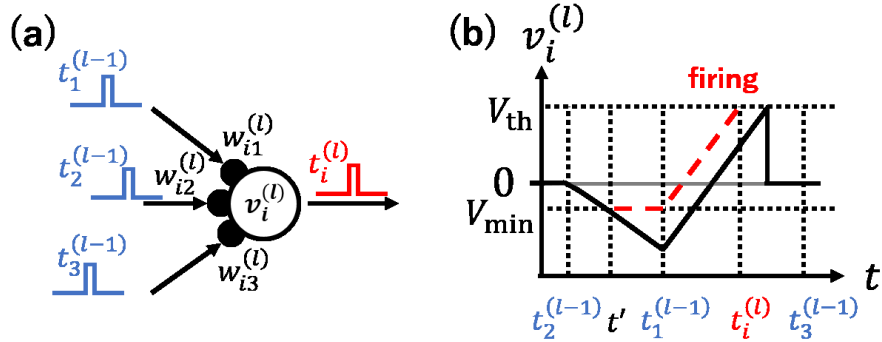}
\caption{(a) Schematic diagram of the $i$th  neuron in the $l$th layer receiving spikes from neurons in the $l-1$th layer at $t_1^{(l-1)}$, $t_2^{(l-1)}$, $t_3^{(l-1)}$, and firing at $t_i^{(l)}$ . 
(b) Schematic diagram of time evolution of the membrane potential of the $i$ th neuron in the $l$th layer 
\textcolor{\mycolor}{ with (red dashed line) and without (black solid line) a lower bound of the membrane potential. The upper and lower horizontal black dotted lines represent the values of $V_\text{th}$ and $V_\text{min}$, respectively.}
}
\label{fig:model}
\end{center}
\end{figure}

We adopted a simple integrate-and-fire neuron model in a multilayer network that was suitable for VLSI implementation \cite{Maass1999}\cite{Sakemi2020supervised}:
\begin{flalign}
\frac{d}{dt} v_{i}^{(l)} (t) &= \sum _j ^N w_{ij}^{(l)} \theta(t- t_j^{(l-1)})  \label{eq:neuron}\\
\theta (x) &= \begin{cases}
0,~(x<0)\\
1,~(x\ge0)
\end{cases}
\end{flalign}
where $v_{i}^{(l)}\textcolor{\mycolorSec}{(t)}$ is the membrane potential of the $i$th neuron in the $l$th layer, $w_{ij}^{(l)}$ is the weight from the $j$th neuron in the $l-1$th layer to the $i$th neuron in the $l$th layer, and $t_j^{(l-1)}$ is the spike timing of the $j$th neuron in the $l-1$th layer.
When the membrane potential exceeds the firing threshold $V_\text{th}$, the neuron generates a spike.
\textcolor{\mycolorSec}{Then, the membrane potential is set to $0$ and the potential does not change again.}
Fig. \ref{fig:model} shows schematic diagram of the neuron model.

For image processing in SNNs, the normalized intensity of the $i$th pixel of an image $x_i \in (0,1]$ 
is converted into an input spike, with the timing given by 
\begin{flalign}
t_i^{(0)} = \tau ( 1 - x_i)  \label{eq:input_coding}
\end{flalign}
where $\tau$ is the time span of the input spikes and it is set to 5 ms.
When $x_i=0$, the $i$th input spike is not generated. 
To improve the generalization ability, Gaussian noise was introduced into the timing of the input spikes in the training phase. 
\textcolor{\mycolor}{The SNNs were trained with the backpropagation algorithm described in \cite{Sakemi2020supervised}.}

We investigated the following effects in the VLSI implementation of SNNs: 
(i) the discretization of the spike timing of the neurons, 
(ii) the quantization of the weights,
(iii) the lower bound of the membrane potentials, and 
(iv) the \textcolor{\mycolorSec}{temporal fluctuation of} the firing threshold. 
In the remainder of this section, we explain the experimental setup used to investigate these effects.
We note that training is performed offline, and only the inference is carried out on a VLSI circuit.
Therefore, we did not consider effects (i) to (iv) when training the SNNs. 

After training the SNNs, we discretized the timing of the spikes as follows:
\begin{flalign}
t_i^{(l)} &:= T_\text{clock}^{(\text{model})} t_i^{(l,\text{step})} \\
t_i^{(l,\text{step})} &:= \text{min} \{p \in Z |  v_i^{(l)}(T_\text{clock}^{(\text{model})} p) \ge V_\text{th}^\text{\textcolor{\mycolorSec}{(model)}} \} \label{eq:t_step}
\end{flalign}
where $T_\text{clock}^{(\text{model})}$ is a positive real number representing the time step of the spike timing and $t_i^{(l,\text{step})}$ is an integer.
The time-discretized spikes are sent to neurons in the subsequent layer.
The synaptic weights $w_{ij}^{(l)}$ are replaced with the quantized weights $w^{(\text{min})} w_{ij}^{(l, \text{level})}$, 
where $w^{(\text{min})}$ is a positive real number and $w_{ij}^{(\text{level})}$ is an integer given by $round\left(w_{ij}^{(l)}/w^{(\text{min})}\right)$.


The membrane potential provided by (\ref{eq:neuron}) can take any value less than $V_\text{th}$. 
However, the range of the membrane potential is limited in analog \textcolor{\mycolor}{VLSI} circuits.
To investigate the effects of \textcolor{\mycolor}{such a limited} range, 
we modified the neuron equation (\ref{eq:neuron}) so that the membrane potential could not take a value less than $V_\text{min}$:
\begin{flalign}
\frac{d}{dt} v_{ij}^{(l)} (t) =\begin{cases}
0, \text{ if } v_{ij}^{(l)} (t) = V_\text{min} \text{ and } \sum _j ^N w_{ij}^{(l)} \theta(t- t_j^{(l-1)}) < 0 \\
\sum _j ^N w_{ij}^{(l)} \theta(t- t_j^{(l-1)}), \text{ otherwise.} \\
\end{cases} \label{eq:modified_neuron}
\end{flalign}
Fig. \ref{fig:model} (b) presents a schematic of the evolution of the membrane potential according to (\ref{eq:modified_neuron}).
The solid line represents the membrane potential when the lower bound of the membrane potential is not considered.
The dashed line represents the membrane potential when the lower bound of the membrane potential is $V_\text{min}$. 
In the latter case, the membrane potential is $V_\text{min}$ from time $t'$ to time \textcolor{\mycolor}{$t_1^{(l-1)}$}, where the sum of the inputs $\sum _j ^N w_{ij}^{(l)} \theta(t- t_j^{(l-1)})$ is less than 0.
For $t>t_1^{(l-1)}$, where the sum of the inputs is positive, the membrane potential begins to increase. 
As a result, the spike timing $t_i^{(l)}$ when the lower bound is considered is different from the spike timing  when the lower bound is not considered.
We note that when $V_\text{min}$ is sufficiently low, the membrane potentials obtained by (\ref{eq:modified_neuron}) converge to the membrane potentials obtained by (\ref{eq:neuron}). 

Finally, to investigate the effects of the \textcolor{\mycolorSec}{temporal fluctuation of} the firing threshold, 
we modeled the firing threshold as a time-dependent function $V_\text{th}(t)$.
The value of $V_\text{th}(t)$ is drawn from the Gaussian distribution as 
\begin{flalign}
V_\text{th}(i) \sim N(V_\text{th}, \sigma _{V_\text{th}}),~i=0,1,\dots
\end{flalign}
where $N(a,b)$ is a Gaussian distribution with a mean of $a$ and standard deviation of $b$.
The standard deviation of the firing threshold is given by $\sigma _{V_\text{th}}$.

\subsection{Results}


\textcolor{\mycolor}{ In this section, we demonstrate the effects of VLSI circuit constraints on SNNs (784-800-10) for the MNIST \cite{Lecun1998gradient} and Fashion-MNIST \cite{Xiao2017fashion} datasets.}


\begin{figure}
\begin{center}
\includegraphics[clip, width=12cm]{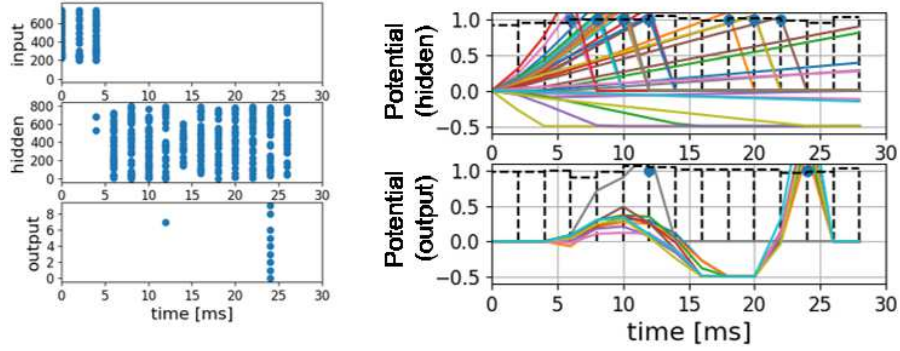}
\caption{Typical results of raster plots (left panels) and membrane potentials (right panels) for MNIST dataset. $T_\text{clock}^\text{(model)}$ = 2 ms, $V_\text{min}$ = -0.5, and $\sigma _{V_\text{th}}$ = 0.04. Weights are not quantized. 
The horizontal dashed lines in the right panels represent the values of the \textcolor{\mycolorSec}{temporally fluctuating} firing threshold $V_\text{th}$.}
\label{fig:raster_discretized_MNIST}
\end{center}
\end{figure}

Fig. \ref{fig:raster_discretized_MNIST} presents the timing of the spikes (raster plots) and time evolution of the membrane potentials for the MNIST dataset  
when $T_\text{clock}^\text{(model)}$ = 2 ms, $V_\text{min}$ = -0.5, and  $\sigma_{V_\text{th}}$ = 0.04.
\textcolor{\mycolor}{In the raster plot,} the timing of the spikes lay only on integer numbers multiplied by $T_\text{clock}^\text{(model)}$. 
The slope of the membrane potential changed only at time discretized by $T_\text{clock}^\text{(model)}$.
Moreover, the membrane potential never took values lower than $V_\text{min}$.

\begin{figure}
\begin{center}
\includegraphics[clip, width=12cm]{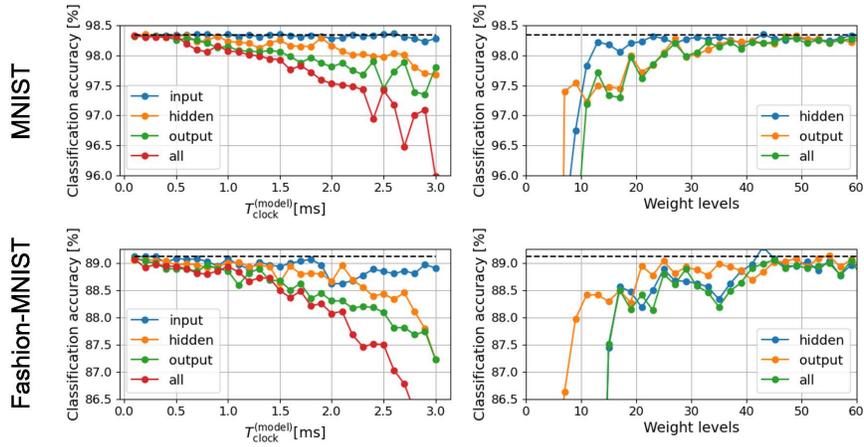}
\caption{Classification accuracy of SNNs on MNIST and Fashion-MNIST datasets \textcolor{\mycolor}{with time-discretization (left panels) and weight quantization (right panels)}.
The horizontal dashed lines in \textcolor{\mycolor}{all panels} represent the performance when no circuit constraints are considered.
}
\label{fig:time_discretized_MNIST}
\end{center}
\end{figure}

\textcolor{\mycolor}{The left panels of} Fig. \ref{fig:time_discretized_MNIST} depicts the performance of the SNNs with time-discretized spikes for various values of $T_\text{clock}^\text{(model)}$.
For both the MNIST and Fashion-MNIST datasets, the effects of the time discretization of the spikes at the output layer were more significant than those at the other layers.
This can be attributed to the fact that 
the prediction results obtained by the network are represented by the earliest spike at the output layer. 
If the real timing difference between the earliest and second earliest spikes is less than $T_\text{clock}^\text{(model)}$, 
the network cannot predict the correct label when the timing of the spikes is discretized.
The effects of the time discretization of the spike timing were substantially lower for the input layer.
This may be because the coding of the input spikes includes ``no spike'' states, which are inherently robust against timing discretization.


\textcolor{\mycolor}{The right panels of} Fig. \ref{fig:time_discretized_MNIST} presents the performance of the SNNs when the weights were quantized. 
In this figure, the horizontal axis represents the number of levels of the quantized weights, given by $\text{max }\left( |w^{(l)}_{ij}|\right) / w^\text{(min)}$.
We found that performance degradation could be avoided if the number of levels was greater than approximately 20. 

\begin{figure}
\begin{center}
\includegraphics[clip, width=12cm]{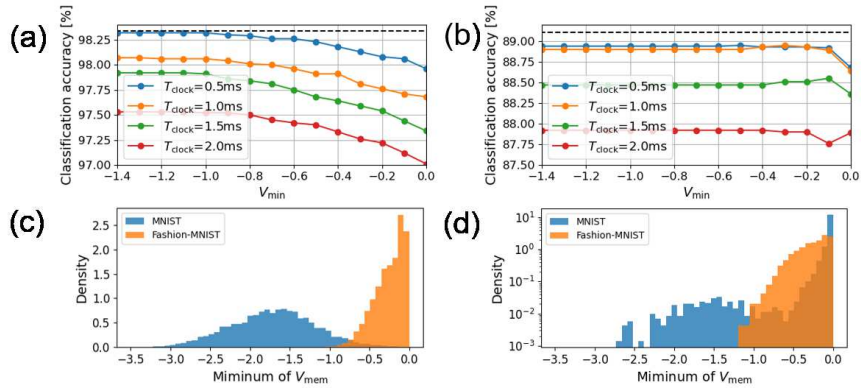}
\caption{Classification accuracy for various values of lower bound $V_\text{min}$ of membrane potential for (a) MNIST dataset and (b) Fashion-MNIST dataset. The distributions of the lowest membrane potentials in  the output layer (c) and those of the lowest membrane potential before the earliest timing of spikes in the output layer (d) when $T_\text{clock}^\text{(model)}$ = 0.6 ms for the above datasets. 
The horizontal dashed lines in (a) and (b) represent the performance when no circuit constraints are considered.
}
\label{fig:Vmin}
\end{center}
\end{figure}

Figs. \ref{fig:Vmin} (a) and (b) present the classification accuracy for various values of the lower bound $V_\text{min}$ of the membrane potential.
Fig. \ref{fig:Vmin} (c) shows the distribution of the minimum value of the membrane potential of the neurons at the output layer, 
given by 
\begin{flalign}
v^\text{(min)}:=\text{min}_{i}\left( \text{min}_{t < t_i^{(2)}} v_i^\text{(2)}(t)\right) \label{eq:v_min}
\end{flalign}
when $V_\text{min}$ was not considered. 
Because the membrane potentials were more likely to take lower values for the MNIST dataset than for the Fashion-MNIST dataset,
the SNNs trained on the MNIST dataset were more vulnerable to the value of $V_\text{min}$ \textcolor{\mycolor}{as indicated in Figs. \ref{fig:Vmin} (a) and (b)}.
We note that the performance was not affected when only the lower bound the membrane potential of neurons in the hidden layer were considered.

One may wonder why the SNNs trained on the MNIST dataset were almost not affected by the lower bound when $V_\text{min}$ = -1, 
even though the average of the minimum voltage $v^\text{(min)}$ given by (\ref{eq:v_min}) was much lower than -1 as shown in Fig. \ref{fig:Vmin} (c). 
This is because the minimum values (\ref{eq:v_min}) were typically obtained after another neuron was fired, as shown in Fig. \ref{fig:raster_discretized_MNIST}.
To elucidate this fact, Fig. \ref{fig:Vmin} (d) plots the distribution of the minimum membrane potential before the timing of the earliest spike:
\begin{flalign}
\hat{v}^\text{(min)}:=\text{min}_{i} \left(\text{min}_{t < \text{min}_j t_j^{(2)}}v_i^\text{(2)}(t)\right). \label{eq:v_min_2}
\end{flalign}
We found that a large portion of the minimum membrane potentials given by (\ref{eq:v_min_2}) had distributions that were higher than -0.5. 

\begin{figure}
\begin{center}
\includegraphics[clip, width=12.5cm]{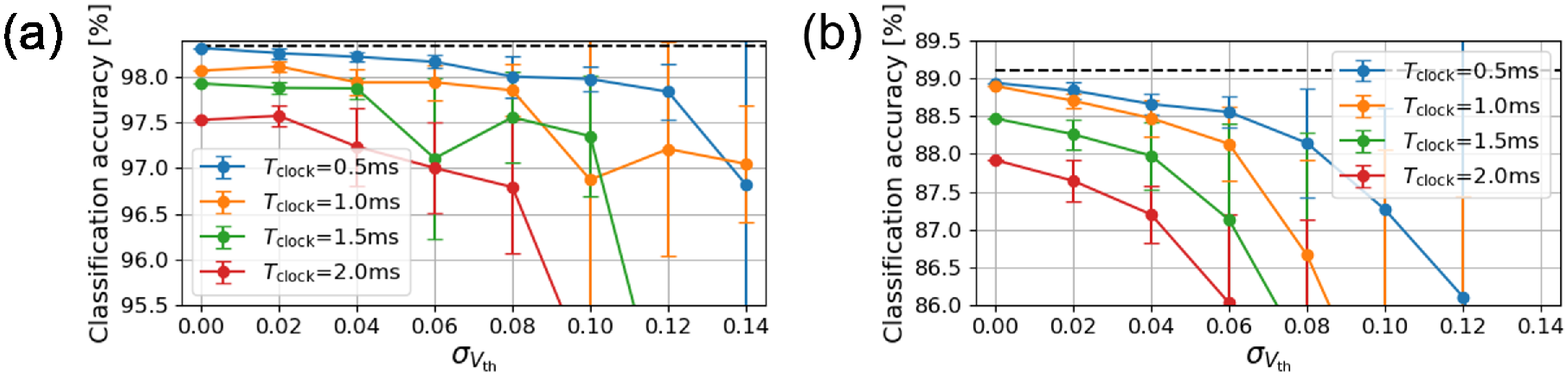}
\caption{Classification accuracy for various values of $\sigma_{V_\text{th}}$ for (a) MNIST dataset and (b) Fashion-MNIST dataset.
The horizontal dashed lines in (a) and (b) represent the performance when any circuit constraints are not considered.
}
\label{fig:Vth_variation}
\end{center}
\end{figure}

Fig. \ref{fig:Vth_variation} depicts the classification accuracy for various values of $\sigma _{V_\text{th}}$.
We found that the performance of the SNNs was not significantly impaired if $\sigma _{V_\text{th}}$ was less than 0.05 for the MNIST  and Fashion-MNIST datasets.

\section{Mapping of SNNs to \textcolor{\mycolor}{VLSI} circuits} \label{sec:mapping}

We demonstrate a procedure for mapping SNNs to VLSI circuits.
First, we introduce a VLSI circuit model derived from Kirchhoff's law.
Second, we show how to map an SNN to the VLSI circuit optimally, assuming certain circuit parameters and constraints.

\subsection{Circuit model}

\begin{figure}
\begin{center}
\includegraphics[clip, width=11cm]{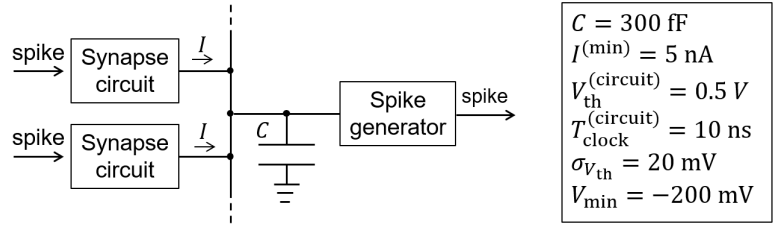}
\caption{\textcolor{\mycolor}{Schematic of} integrate-and-fire neuron circuit. 
Currents from many synapses are converted into voltage at the capacitor $C$. 
A spike is generated when the voltage exceeds a threshold.}
\label{fig:neuron_circuit}
\end{center}
\end{figure}

We consider an integrate-and-fire model as illustrated in Fig. \ref{fig:neuron_circuit}, which is commonly employed in matrix--vector computation \cite{Wang2016}\cite{Wang2018time}\cite{Bavandpour2019}\cite{Yamaguchi2021energy} and SNNs \cite{Sakemi2020supervised,Oh2020hardware}.
The time evolution of the membrane potential represented by the voltage at a capacitor is obtained by Kirchhoff's current law, as follows:
\begin{flalign}
&C \frac{d}{dt} v_i^{(l)} (t) 
= I^\text{(min)} \sum _j ^N I_{ij}^\text{(level)} \theta(t- t_j^{(l-1)}) \label{eq:neuron_circuit}
\end{flalign}
where $I^\text{(min)}$ represents the controllable minimum current, $I_{ij}^\text{(level)}$ is an integer, and $C$ is the capacitance.
The timing of the firing is obtained as follows:
\begin{flalign}
t_i^{(l)} &:= T_\text{clock}^{(\text{circuit})} t_i^{(l,\text{step})} \\
t_i^{(l,\text{step})} &= \text{min} \{p \in Z |  v_i^{(l)}(T_\text{clock}^{(\text{circuit})} p) \ge V_\text{th}^\text{\textcolor{\mycolorSec}{(circuit)}} \} \label{eq:t_step_circuit}
\end{flalign}
where $T_\text{clock}^{(\text{circuit})}$ is a real number representing the time step of the time-discrete spikes in the circuits.
%

By setting $I_{ij}^{(l, \text{level})} = w_{ij}^{(l, \text{level})}$, 
the condition whereby \textcolor{\mycolor}{(\ref{eq:neuron}) and (\ref{eq:neuron_circuit})} are equivalent can be obtained as follows:
\begin{flalign}
T_\text{clock} ^\text{(model)} w^\text{(min)} = \frac{T_\text{clock} ^\text{(circuit)}I^\text{(min)} \textcolor{\mycolorSec}{V_\text{th}^\text{(model)}}}{C \textcolor{\mycolorSec}{V_\text{th}^\text{(circuit)}}}.   \label{eq:mapping_constraint}
\end{flalign}

\subsection{Optimal mapping of SNNs to circuits}


We used the circuit parameters displayed in \textcolor{\mycolor}{Fig. \ref{fig:neuron_circuit}} to study the mapping of SNNs to circuits.
These values of parameters were determined typically from VLSI design based on the approach of time-domain analog computing with transient states (TACT), in which synapse currents were given by field-effect transistors operating in a subthreshold region~\cite{Wang2018time,Yamaguchi2021energy}.


\begin{figure}
\begin{center}
\includegraphics[clip, width=12cm]{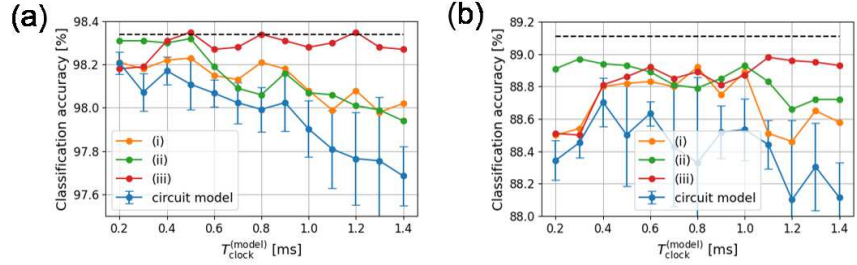}
\caption{Classification accuracy on (a) MNIST dataset and (b) Fashion-MNIST dataset with circuit model for various values of $T_\text{clock}^\text{(model)}$. 
Classification accuracy is depicted when the circuit model is considered, (i) when only the effects of time discretization and weight quantization are considered, (ii) when only the effects of time discretization is considered, and (iii) when only the effects of weight quantization is considered. Note that the effects of the lower bound of the membrane potential and \textcolor{\mycolorSec}{the temporal fluctuation of} firing threshold are not included in (i)-(iii).
The horizontal dashed lines in (a) and (b) represent the performance when no circuit constraints are considered.
}
\label{fig:MNIST_circuit}
\end{center}
\end{figure}

Fig. \ref{fig:MNIST_circuit} presents the classification accuracy for various values of $T_\text{clock}^\text{(model)}$.
Note that the value of $w^\text{(min)}$ was determined using the values of $T_\text{clock}^\text{(model)}$ and (\ref{eq:mapping_constraint}).
For the MNIST dataset, the classification accuracy decreased almost monotonically as $T_\text{clock}^\text{(model)}$ increased.
For the Fashion-MNIST dataset, the classification accuracy exhibited a peak around $T_\text{clock}^\text{(model)}$ = \textcolor{\mycolorSec}{0.}4 ms. 
The performance of the SNN was mainly affected by the time discretization and weight quantization. 
To elucidate that, Fig. \ref{fig:MNIST_circuit} also depicts the classification accuracy 
when only the time discretization and the weight quantization were considered and that 
when only the time discretization or the weight quantization was considered.

\begin{figure}[b]
\begin{center}
\includegraphics[clip, width=7cm]{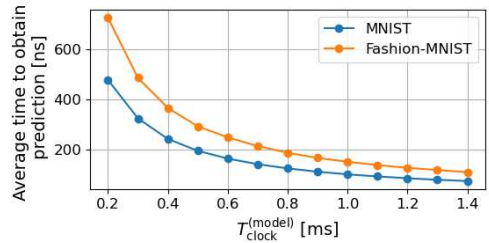}
\caption{Average time of earliest spike at output layer for various values of $T_\text{clock}^\text{(model)}$. }
\label{fig:prediction_time}
\end{center}
\end{figure}

As the network could cease its operation after the earliest spike was generated in the output layer,
the energy efficiency increased as the average timing of the earliest spike became shorter.
Fig. \ref{fig:prediction_time} indicates that the average timing of the earliest spike at the output layer decreased as a function of $T_\text{clock}^\text{(model)}$, because the time steps required to obtain the output spikes was small when  $T_\text{clock}^\text{(model)}$ was large (see Fig. \ref{fig:raster_discretized_MNIST}) with the same $T_\text{clock}^\text{(circuit)}$. 

The optimal choice of $T_\text{clock}^\text{(model)}$ was determined by the trade-off between the classification accuracy (Fig. \ref{fig:MNIST_circuit}) and energy efficiency (Fig.  \ref{fig:prediction_time}).
For example, if an accuracy of more than 98\% was required with the minimum energy efficiency for the MNIST dataset, the optimal $T_\text{clock}^\text{(model)}$ would be approximately 0.8 ms.

\section{Conclusions}

In this study, we evaluated the performance of SNNs with TTFS coding on the MNIST and Fashion-MNIST datasets 
when considering circuit constraints. 
We found that the SNNs with TTFS coding were robust against these circuit constraints. 
Although information is contained in the precise spike timing in SNNs, 
the performance of the SNNs was only slowly degraded as the time step of the time discretization increased. 
Only 20 or 30 quantized levels were sufficient for the weights to prevent performance impairment, which corresponded to 4- or 5-bit precision.
The lower bound of the membrane potential was not critical if we used only the earliest spike as the prediction result.
The \textcolor{\mycolorSec}{temporal fluctuation of} the firing threshold was also not critical. 

We demonstrated the optimal mapping of SNN models with TTFS coding to VLSI circuits, assuming a reasonable set of circuit parameters.
We confirmed that the optimal choice of the discretization parameters for the spike timing and weights is obtained by a trade-off between the classification accuracy and energy efficiency. 
In this study, we fixed the circuit parameters and searched for the optimal discretization parameters for the SNNs.
It will be straightforward to expand this strategy so that one can also search for optimal circuit parameters such as the clock period $T_\text{clock}^\text{(circuit)}$ and the capacitance of neurons $C$.
\textcolor{\mycolorSec}{Further, we focused on a specific neuron model (\ref{eq:neuron}), however, the proposed techniques can be applied to other neuron models such as ones adopted in \cite{Mostafa2018supervised} and \cite{Comsa2019temporal}.}

\section*{Acknowledgment}
The authors would like to thank Kazumasa Yanagisawa and  Masatoshi Yamaguchi from Floadia corporation for the fruitful discussion.

%
%
%
\bibliographystyle{splncs04}
\bibliography{myBib}

\begin{thebibliography}{10}
\providecommand{\url}[1]{\texttt{#1}}
\providecommand{\urlprefix}{URL }
\providecommand{\doi}[1]{https://doi.org/#1}

\bibitem{Bavandpour2019}
Bavandpour, M., Mahmoodi, M.R., Strukov, D.B.: Energy-efficient time-domain
  vector-by-matrix multiplier for neurocomputing and beyond. IEEE Transactions
  on Circuits and Systems II: Express Briefs  (2019)

\bibitem{Bohte2002error}
Bohte, S.M., Kok, J.N., Poutr^^c3^^a9, H.L.: Error-backpropagation in
  temporally encoded networks of spiking neurons. Neurocomputing
  \textbf{48}(1),  17--37 (October 2002)

\bibitem{Comsa2019temporal}
Comsa, I.M., et~al.: Temporal coding in spiking neural networks with alpha
  synaptic function. In: ICASSP 2020 - 2020 IEEE International Conference on
  Acoustics, Speech and Signal Processing (ICASSP). pp. 8529--8533 (2020)

\bibitem{Davies2018loihi}
Davies, M., et~al.: Loihi: A neuromorphic manycore processor with on-chip
  learning. IEEE Micro  \textbf{38}(1),  82--99 (2018)

\bibitem{Diehl2015fast}
{Diehl}, P.U., et~al.: Fast-classifying, high-accuracy spiking deep networks
  through weight and threshold balancing. In: 2015 International Joint
  Conference on Neural Networks (IJCNN). pp.~1--8 (July 2015)

\bibitem{Goltz2020fast}
G{\"o}ltz, J., et~al.: Fast and deep neuromorphic learning with first-spike
  coding. In: Proceedings of the Neuro-inspired Computational Elements
  Workshop. pp.~1--3 (2020)

\bibitem{Huh2018gradient}
Huh, D., Sejnowski, T.J.: Gradient descent for spiking neural networks. In:
  Advances in Neural Information Processing Systems. vol.~31, pp. 1433--1443
  (2018)

\bibitem{Kheradpisheh2019s4nn}
Kheradpisheh, S.R., Masquelier, T.: Temporal backpropagation for spiking neural
  networks with one spike per neuron. International Journal of Neural Systems
  \textbf{30}(6),  2050027 (2020)

\bibitem{Kheradpisheh2020bs4nn}
Kheradpisheh, S.R., Mirsadeghi, M., Masquelier, T.: {BS4NN}: Binarized spiking
  neural networks with temporal coding and learning. arXiv:2007.04039  (2020)

\bibitem{Kim2020unifying}
Kim, J., Kim, K., Kim, J.J.: Unifying activation-and timing-based learning
  rules for spiking neural networks. arXiv:2006.02642  (2020)

\bibitem{Kim2020spiking}
Kim, S., Park, S., Na, B., Yoon, S.: Spiking-{YOLO}: spiking neural network for
  energy-efficient object detection. In: Proceedings of the AAAI Conference on
  Artificial Intelligence. vol.~34, pp. 11270--11277 (2020)

\bibitem{Lecun1998gradient}
{LeCun}, Y., {Bottou}, L., {Bengio}, Y., {Haffner}, P.: Gradient-based learning
  applied to document recognition. Proceedings of the IEEE  \textbf{86}(11),
  2278--2324 (November 1998)

\bibitem{Maass1999}
Maass, W.: ``Computing with Spiking Neurons," \textrm{in} Pulsed Neural
  Networks, chap.~2, pp. 55--85. MIT Press (1999)

\bibitem{Merolla2014}
Merolla, P.A., et~al.: A million spiking-neuron integrated circuit with a
  scalable communication network and interface. Science  \textbf{345}(6197),
  668--673 (2014)

\bibitem{Moradi2018scalable}
{Moradi}, S., {Qiao}, N., {Stefanini}, F., {Indiveri}, G.: A scalable multicore
  architecture with heterogeneous memory structures for dynamic neuromorphic
  asynchronous processors ({DYNAP}s). IEEE Transactions on Biomedical Circuits
  and Systems  \textbf{12}(1),  106--122 (Feb 2018)

\bibitem{Mostafa2018supervised}
{Mostafa}, H.: Supervised learning based on temporal coding in spiking neural
  networks. IEEE Transactions on Neural Networks and Learning Systems
  \textbf{29}(7),  3227--3235 (July 2018)

\bibitem{Oh2020hardware}
Oh, S., et~al.: Hardware implementation of spiking neural networks using
  time-to-first-spike encoding. arXiv:2006.05033  (2020)

\bibitem{Pfeiffer2018Deep}
Pfeiffer, M., Pfeil, T.: Deep learning with spiking neurons: Opportunities and
  challenges. Frontiers in Neuroscience  \textbf{12}(774),  1--18 (2018)

\bibitem{Sakemi2020supervised}
Sakemi, Y., Morino, K., Morie, T., Aihara, K.: A supervised learning algorithm
  for multilayer spiking neural networks based on temporal coding toward
  energy-efficient {VLSI} processor design. arXiv:2001.05348  (2020)

\bibitem{Sze2017}
{Sze}, V., {Chen}, Y., {Yang}, T., {Emer}, J.S.: Efficient processing of deep
  neural networks: A tutorial and survey. Proceedings of the IEEE
  \textbf{105}(12),  2295--2329 (2017)

\bibitem{Tavanaei2019Deep}
Tavanaei, A., Ghodrati, M., Kheradpisheh, S.R., Masquelier, T., Maida, A.: Deep
  learning in spiking neural networks. Neural Networks  \textbf{111},  47--63
  (2019)

\bibitem{Wang2018time}
Wang, Q., Tamukoh, H., {Morie}, T.: A time-domain analog weighted-sum
  calculation model for extremely low power {VLSI} implementation of
  multi-layer neural networks. arXiv:1810.06819  (2018)

\bibitem{Wang2016}
Wang, Q., Tamukoh, H., Morie, T.: Time-domain weighted-sum calculation for
  ultimately low power {VLSI} neural networks. In: Neural Information
  Processing. vol.~9947, pp. 240--247 (September 2016)

\bibitem{Xiao2017fashion}
Xiao, H., Rasul, K., Vollgraf, R.: Fashion-{MNIST}: a novel image dataset for
  benchmarking machine learning algorithms. arXiv:1708.07747  (2017)

\bibitem{Yamaguchi2021energy}
{Yamaguchi}, M., {Iwamoto}, G., {Nishimura}, Y., {Tamukoh}, H., {Morie}, T.: An
  energy-efficient time-domain analog cmos binaryconnect neural network
  processor based on a pulse-width modulation approach. IEEE Access
  \textbf{9},  2644--2654 (2020)

\end{thebibliography}


%
%
%
%
%
\end{document}